# Tilted and type-III Dirac cones emerging from flat bands in photonic orbital graphene

M. Milicevic[1], G. Montambaux[2], T. Ozawa[3], I. Sagnes[1], A. Lemaître[1], L. Le Gratiet[1], A. Harouri[1], J. Bloch[1], A. Amo[4*]

[1]*Centre de Nanosciences et de Nanotechnologies, CNRS, Univ. Paris-Sud, Université Paris-Saclay, C2N-Marcoussis, 91460 Marcoussis, France*

[2]*Laboratoire de Physique des Solides, CNRS, Univ. Paris-Sud, Université Paris-Saclay, 91405 Orsay Cedex, France*

[3] *Interdisciplinary Theoretical and Mathematical Sciences Program (iTHEMS), RIKEN, Wako, Saitama 351-0198, Japan*

[4]*Univ. Lille, CNRS, UMR 8523 – PhLAM – Physique des Lasers Atomes et Molécules, F-59000 Lille, France*

[*]alberto.amo-garcia@univ-lille.fr

**The extraordinary electronic properties of Dirac materials, the two-dimensional partners of Weyl semimetals, arise from the linear crossings in their band structure. When the dispersion around the Dirac points is tilted, the emergence of intricate transport phenomena has been predicted, such as modified Klein tunnelling, intrinsic anomalous Hall effects and ferrimagnetism. However, Dirac materials are rare, particularly with tilted Dirac cones. Recently, artificial materials whose building blocks present orbital degrees of freedom have appeared as promising candidates for the engineering of exotic Dirac dispersions. Here we take advantage of the orbital structure of photonic resonators arranged in a honeycomb lattice to implement photonic lattices with semi-Dirac, tilted and, most interestingly, type-III Dirac cones that combine flat and linear dispersions. The tilted cones emerge from the touching of a flat and a parabolic band with a non-trivial topological charge. These results open the way to the synthesis of orbital Dirac matter with unconventional transport properties and, in combination with polariton nonlinearities, to the study of topological and Dirac superfluids in photonic lattices.**

The extraordinary transport properties of Dirac materials arise from the spinor nature of their electronic wavefunctions and from the linear dispersion around Dirac and Weyl points. In two-dimensions, Klein tunnelling, weak antilocalisation, unconventional Landau levels or bulk pseudo-confinement appear as some of their most remarkable features[1,2]. Standard Dirac cones, like those present in graphene and other two-dimensional materials, have rotational symmetry about the Dirac quasi-momentum. Their topological properties make them particularly robust to deformations of the lattice: Dirac cones always appear in pairs, each of them characterised by a topological charge[3], and in the presence of time-reversal and inversion symmetry, they can only be annihilated via their merging with a Dirac point of opposite charge[4–9].

Dirac cones can be classified according to the geometry of their Fermi surface. The cylindrically symmetric Dirac cones described above belong to the family of type-I Dirac cones. They are characterised by a closed Fermi surface eventually becoming a single point at the band crossing, where the density of states vanishes (Fig. 1a). However, they are not the only kind of linear band crossings that can be found in Dirac materials. The general Hamiltonian describing a Dirac cone in two dimensions can be expressed as[10] $H(\boldsymbol{q}) = (v_{0x}q_x + v_{0y}q_y)\sigma_0 + v_x q_x \sigma_a + v_y q_y \sigma_b$, where $q_{x,y}$ is the wave vector measured from the Dirac point; $v_{0x}, v_{0y}, v_x$ and $v_y$ represent effective velocities; $\sigma_0$ is the $2 \times 2$ identity matrix and $\sigma_{a,b} = \boldsymbol{u}_{a,b} \cdot \boldsymbol{\sigma}$, where $\boldsymbol{u}_{a,b}$ are suitably chosen orthogonal unit vectors and $\boldsymbol{\sigma} = (\sigma_x, \sigma_y, \sigma_z)$ is the vector of Pauli matrices. The eigenenergies of this Hamiltonian form two bands:

$$E_\pm(\boldsymbol{k}) = v_{0x}q_x + v_{0y}q_y \pm \sqrt{(v_x q_x)^2 + (v_y q_y)^2}. \qquad (1)$$

If both coefficients $v_{0x}$ and $v_{0y}$ are equal to zero we obtain the energy spectrum of a type-I Dirac cone with Fermi velocities $v_x$ and $v_y$. If any of the $v_{0x}, v_{0y}$ coefficients is non zero, then the Dirac cone is tilted (Fig. 1(b)). In two dimensional materials, this kind of tilted Dirac dispersion has been predicted to appear in quinoid-type[10] and hydrogenated graphene[11], and it has been indirectly evidenced in the organic semiconductor[10,12,13] α-(BEDT-TTF)$_2$I$_3$. The interest of materials presenting this dispersion resides in its non-isotropic transport properties, which can be used for valley filtering in p–n junctions[14] or for the generation of photocurrent[15].

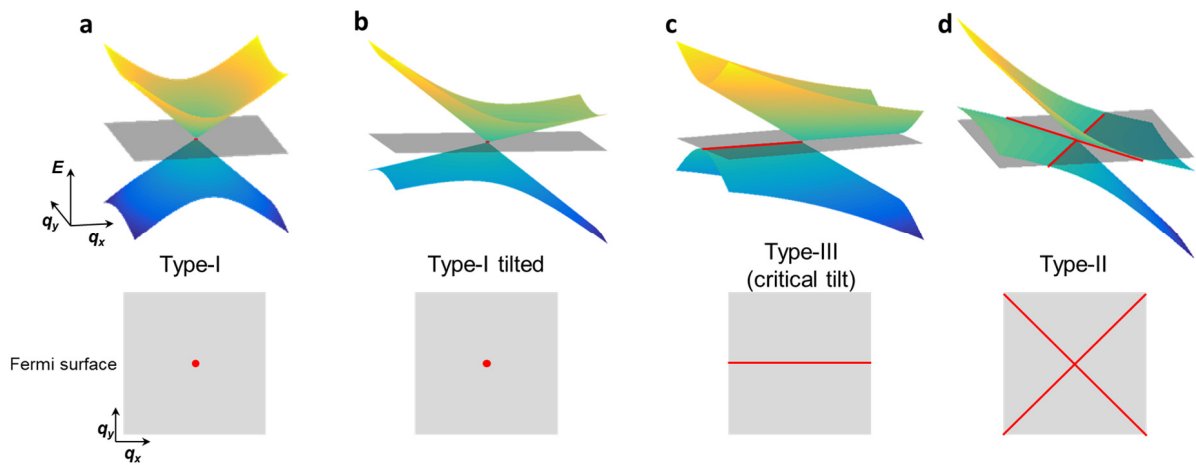

**Figure 1. Types of Dirac dispersions in two dimensions.** (top) Dispersions together with the zero energy plane (grey); (bottom, in red) zero-energy Fermi surface: **a** Standard, type-I Dirac cones characterised by a linear dispersion in all directions in k-space and a point-like Fermi surface. **b** Type-I tilted Dirac cone. **c** Type-III Dirac point (critically tilted), combining a flat band along a line and linear dispersions. Its Fermi surface is a line. **d** Type-II Dirac cone.

When the tilt parameter $\tilde{v}_0 \equiv \sqrt{(v_{0x}/v_x)^2 + (v_{0y}/v_y)^2}$ is larger than 1, a type-II Dirac point[10,16] is formed (Fig. 1d). Its Fermi surface is no longer a point but two crossing lines, and the density of states at the energy of the Dirac point becomes finite[16]. They have been recently observed in the form of Fermi arcs in three-dimensional semimetals[17–19]. A particularly interesting situation takes place at the transition between type-I and type-II Dirac cones, that is, when the tilt parameter $\tilde{v}_0 = 1$. In this case, the cone is critically tilted, with a flat band along one direction (Fig. 1c). Because of its distinct Fermi surface, a single line, and its diverging density of states, this kind of dispersion has been labelled type-III Dirac cone[20,21]. While most of their electric and magnetic properties are still to be unveiled, they have been predicted to greatly enhance the superconducting gap in Weyl semimetals[22], and provide a new platform for the study of correlated phases with a flat band[23].

Type-III Dirac points have not been yet reported experimentally, and tilted type-I and type-II have been challenging to synthetize[13,17,19] because they require materials whose constituent atoms are arranged in lattices with intricate electron hoppings[10,12,16]. Existing proposals rely on the engineering of next-nearest neighbours tunnelling, difficult to find in natural materials and to implement in electronic metamaterials. Artificial photonic lattices represent an opportunity to explore the physics of unconventional Dirac points thanks to the at-will control of onsite energies and hoppings[24]. Current schemes are based on the design of long distant coupling of photons in lattices of resonators[25,26]; evidence of type-II Weyl points has been recently reported in microwave metamaterials[27,28] and via conical diffraction in laser written waveguides with elaborate couplings[29].

In this article we propose and demonstrate experimentally a new method to implement tilted and semi-Dirac cones for photons, and provide the first experimental observation of type-III Dirac cones. We employ $p_x, p_y$ orbital bands in a honeycomb lattice of polariton micropillars, which arise from the nearest-neighbour hopping of photons confined in the first excited modes of each resonator of the lattice[30,31]. In this analogue system, the band structure is directly accessible in photoluminescence experiments. The orbital bands with symmetric hoppings contain a flat band that touches a parabolic band. When asymmetry in the hopping is introduced, which simulates uniaxial strain in solid-state graphene, the flat-parabolic band touching evolves into tilted and type-III Dirac cones. The richness of this multi-band system allows, in addition, the observation of semi-Dirac cones which combine massless and massive dispersions. By analysing their topological charge, i.e. the winding of the Hamiltonian around each Dirac point, we show that the semi-Dirac, tilted and type-III Dirac cones emerge as a consequence of topological Lifshitz transitions induced by strain in the orbital bands. The present realisation shows the potential of orbital bands to engineer the properties of Dirac matter.

**Photonic orbital lattice**

The photonic platform we employ is a honeycomb lattice of coupled micropillars. The lattice is etched from a semiconductor planar microcavity made out of two AlGaAs Bragg mirrors that confine photons in the vertical direction, and twelve GaAs quantum wells embedded in the spacer between the mirrors (see Sec. I of Supplementary Information). At 10 K, the temperature of our experiments, the confined photons and the quantum well excitons are in the strong coupling regime and form polaritons, light-matter hybrid quasi-particles. Each micropillar, with a diameter of 2.75 μm presents an additional lateral confinement due to the index of refraction contrast between the semiconductor and air (Fig. 2a). The photonic spectrum of an individual micropillar is thus discrete with the lowest energy s-mode being cylindrically symmetric, and the first excited modes formed by two degenerate $p_x, p_y$ orbitals with lobes oriented 90º from each other (Fig. 2b)[31]. In the honeycomb lattice, the micropillars overlap (centre-to-centre distance $d = 2.4$ µm) enabling the hopping of photons between adjacent sites (Fig. 2d).

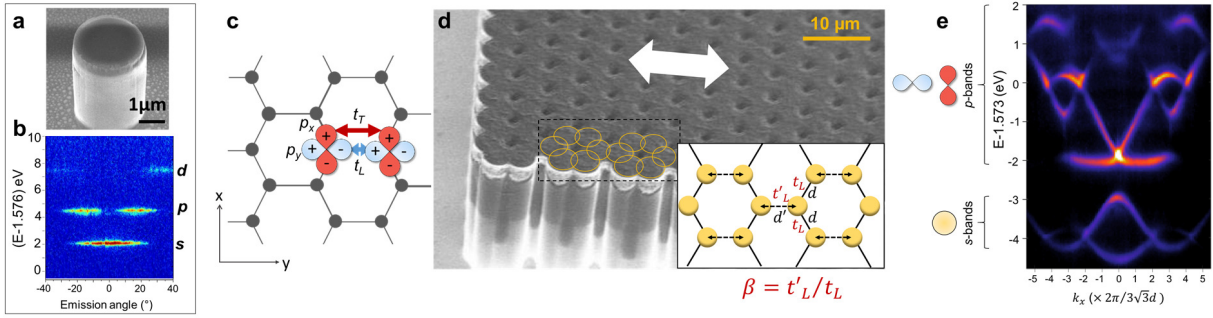

**Figure 2. Honeycomb polariton lattice with orbital bands.** Scanning electron microscopy (SEM) image of a single micropillar, the elementary building block of the honeycomb lattice. **b** Characteristic emission spectrum of a single micropillar, showing s, p and d discrete modes. **c** Scheme of the coupling of $p_x$ and $p_y$ orbitals in the honeycomb lattice with hoppings $t_L \gg t_T$. **d** Image of a honeycomb lattice of micropillars with homogeneous hoppings ($t_L = t'_L$, $\beta = 1$, see inset). The circles pinpoint the upper surface of the micropillars in two adjacent hexagons; the white arrows indicate the direction along which the coupling $t'_L$ is modified. **e** Experimental photoluminescence of the lattice showing s- and p-bands.

The coupling of s-modes results in two bands with a spectrum and eigenmodes very similar to those of electrons in graphene, shown in the low energy part of Fig. 2e and studied in previous works[30,32]. Here we concentrate on the orbital bands which arise from the coupling of p-modes (high-energy set of bands in Fig. 2e). Orbitals oriented along the link between adjacent pillars ($p_y$ in the example of Fig. 2c) present a coupling $t_L$ much stronger than $t_T$, the coupling of orbitals oriented perpendicular to the link ($p_x$ in Fig. 2c).

Figure 3a shows the angle resolved photoluminescence of the p-bands when exciting the lattice at its centre with a non-resonant continuous wave laser at 745 nm, focused on a 3 µm diameter spot. The power of the laser is 6 mW, well below the threshold for any nonlinear effect. The laser creates electrons and holes in the quantum wells that relax down through phonon scattering to form polaritons distributed over the different bands of the lattice. When polaritons recombine via the escape of a photon out of the microcavity, photons are emitted with an in-plane momentum and a frequency that correspond to those of the original polaritons within the lattice of resonators. An angle and energy resolved measurement employing an imaging spectrometer coupled to a CCD allows reconstructing the dispersion relation (see Supplementary Information). To avoid destructive interference effects along high symmetric crystallographic directions, characteristic of bi-partite lattices[33], we record the emission as a function of $k_x$ for $k_y = 4\pi/3d$, passing through the $K$, $\Gamma$, $K'$ points in the second Brillouin zone (dashed line in Fig. 3e). Four bands are observed: the lowest one is flat, while the two central bands present two type-I Dirac points at K and K' and touch the flat band at the $\Gamma$ point[30]. This is in good agreement with a nearest-neighbour tight-binding calculation[34] assuming $t_T = 0$ (white solid lines, see Supplementary Information). In the experiments, the uppermost band deviates from a flat band due to the coupling to higher energy bands (arising from photonic d-orbitals in the micropillars).

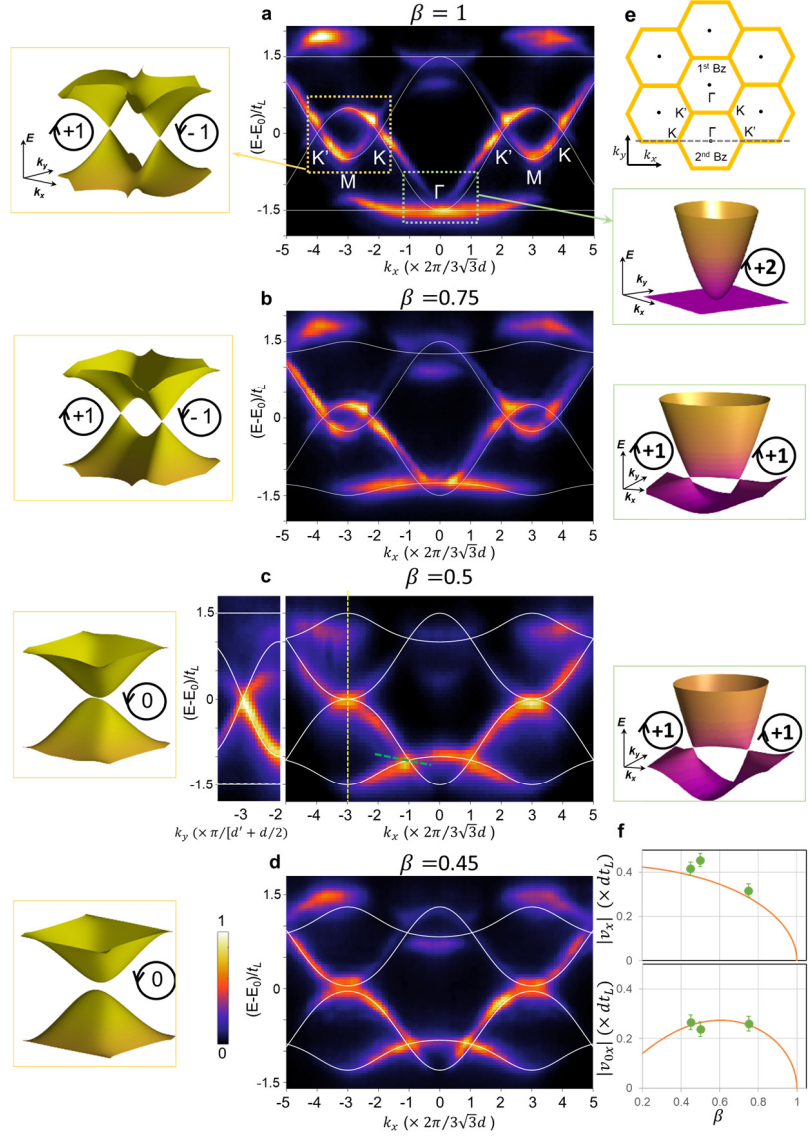

**Figure 3. Tilted Dirac cones in orbital graphene under strain.** The central panels **a**-**d** show the measured polariton photoluminescence intensity as a function of $k_x$ for different values of $\beta$ (the colour scale of each panel has been independently normalised to its maximum value). The cut is done for $k_y = 2\pi/(d' + d/2)$, dashed line in **e**. $d = 2.40$ μm for the unstrained coupling, while for the strained coupling $d'$ is 2.40 μm, 2.60 μm, 2.70 μm and 2.72 μm, respectively, in **a**-**d**. $E_0 = 1.5687$ eV and photon-exciton detuning $\delta = -10$ meV at the energy minimum of the p-bands for panels **a**, **b**, **d**; $E_0 = 1.5780$ meV and $\delta = -2$ meV for **c**. $\hbar t_L = -0.90$ meV for panel **a**, $-0.85$ meV for **b** and **d**, and $-0.65$ for **c**, which are obtained by fitting the measured spectra with a tight-binding Hamiltonian (fits are shown in white lines). The left inset of **c** shows the semi-Dirac dispersion measured along $k_y$ for the value of $k_x$ marked by a dashed line in **c**. The left column, depicts the calculated tight-binding bands for energies close to $E_0$ (orange dashed rectangle in **a**) and the winding of the wavefunctions around the Dirac points. The right column shows the tight-binding bands in the region at the bottom of the bands (green dashed rectangle in **a**) together with the winding of the wavefunctions. **e** Sketch of the Brillouin zones in momentum space. **f** Green dots: absolute value of $v_x$ and $v_{0x}$ extracted from fits of Eq. 1 to the two experimental Dirac cones visible at low energies in **b**-**d** (see Supplementary Information). The plotted value is the average of $|v_x|$ and $|v_{0x}|$ extracted from the two Dirac points. Solid lines: tight-binding result.

## Tilted and semi-Dirac cones

The dispersion of the p-bands can be modified by introducing an artificial uniaxial strain in the lattice. To do so, we change the centre-to-centre distance $d'$ between the micropillars whose link is oriented along the $y$ direction. This is equivalent to modifying $t'_L$, defined in the inset of Fig. 2d, while keeping the other two couplings $t_L$ constant[35]. The central panels of Fig. 3 show the spectra of four different lattices with decreasing strain parameters $\beta \equiv t'_L/t_L$. The value of $\beta$ is extracted from the fit of the tight-binding model (white lines) to the experimental dispersions. Let us first focus on the type-I Dirac cones in the central region (orange rectangle). The left panels depict the calculated dispersion around $E_0$ obtained from the tight-binding model. Decreasing beta, that is, emulating the stretching of the lattice, brings the Dirac cones closer together in the $x$ direction (Fig. 3a-b) until they merge at $\beta = 0.5$ in a single band touching (Fig. 3c). For $\beta < 0.5$ a gap is opened (Fig. 3d and Supplementary Fig. S4). This is a topological Lifshitz transition in which two Dirac cones with opposite topological charge merge and annihilate, predicted for standard s-band graphene[4,36] and reported in photonic[7,8] and atomic[5,6] honeycomb lattices and in black phosphorus[9]. At the merging point ($\beta = 0.5$, Fig. 3c), we provide the direct observation of a semi-Dirac dispersion, with the touching of two parabolic bands along the $k_x$ direction and a linear dispersion along $k_y$ (shown in the left inset of Fig. 3c).

The Dirac points that we have just analysed do not present any tilt: $v_{0x} = v_{0y} = 0$ for any value of the strain. Tilted Dirac cones become apparent when analysing the evolution of the touching between the quadratic and flat bands (green dashed square in Fig. 3a) as a function of strain. For decreasing values of $\beta$, the flat band evolves into a dispersive band with negative effective mass at the $\Gamma$ point, and the band touching divides into two Dirac points that move away from each other in the $k_x$ direction[37] (Fig. 3b-e). Remarkably, they are tilted as indicated by the angle bisector in green dashed lines in Fig. 3c. A fit of Eq. 1 to the experimental dispersions close to the Dirac points reveals the evolution of $v_{0x}$ and $v_x$ as a function of the strain, as depicted in Fig. 3f. The measured values of the tilt ($v_{0x}$) agree well with those expected from the tight-binding Hamiltonian shown as solid lines in Fig. 3f (see Supplementary Information).

## Engineering type-III Dirac cones

Critically tilted type-III Dirac cones with $\tilde{v}_0 = \sqrt{(v_{0x}/v_x)^2 + (v_{0y}/v_y)^2} = 1$ can be engineered in our system when instead of expanding the lattice ($\beta < 1$), it is compressed ($\beta > 1$). This is shown in Fig. 4 for $\beta = 1.5$: in the direction $k_y$ (Fig. 4c), parallel to that along which $d'$ is reduced, two new Dirac points emerge from the flat-parabolic band touching. As a reference, Fig. 4a shows the case of $\beta = 1$ along $k_y$. The most striking feature of the new Dirac points is that they show the crossing of a flat band with zero group velocity and a linear band with finite group velocity (see Fig. 4c, dashed rectangle). This is precisely the signature of a type-III Dirac cone, and it implies $v_{0y} = -v_y$. From the experimental photoluminescence of Fig 4c we measure $|v_{0y}| = (0.29 \pm 0.03)\bar{d}t_L$ and $|v_y| = (0.35 \pm 0.03)\bar{d}t_L$ (with $\bar{d} = \frac{2d'+d}{3}$), which is in good agreement with the tight binding prediction ($|v_{0y}| = |v_y| = 0.37dt_L$, see Supplementary Information). Along the perpendicular direction ($k_x$, Fig. 4d), the type-III cones present a symmetric linear crossing, and we measure $|v_{0x}| = (0.00 \pm 0.04)dt_L$, $|v_x| = (0.46 \pm 0.04)dt_L$, in agreement with the tight-binding Hamiltonian (0 and $0.48dt_L$, respectively), whose dispersion is shown in Fig. 4e. The degeneracy of the flat band results in a divergent density of states at the energy of the type-III Dirac point.

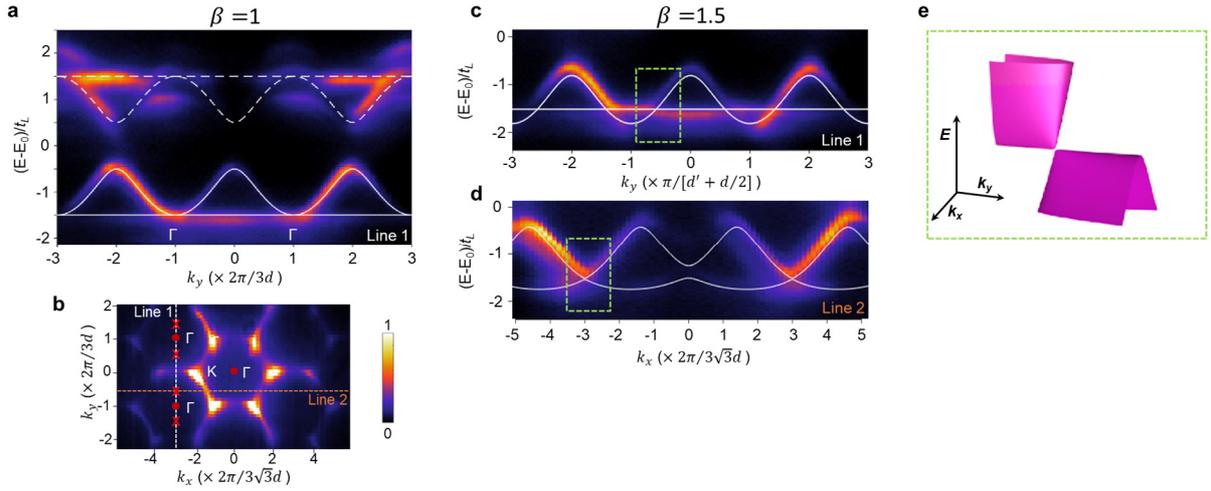

**Figure 4. Type-III Dirac cones. a** Measured photoluminescence intensity for $\beta = 1$ ($d = d' = 2.60$ μm) along $k_y$ for $k_x = -2\pi/\sqrt{3}d$ (line 1 in **b**). $E_0 = 1.5687$ eV and $\hbar t_L = -0.85$ meV. **b** Emission in momentum space at $E_0$ for $\beta = 1$, showing the usual Dirac points at the high symmetry points in reciprocal space. Crosses indicate the position of the Dirac points reported in **c**. **c** Zoom of the low energy section of the measured spectrum for a lattice with $\beta = 1.5$ ($d = 2.60$ μm, $d' = 2.40$ μm) showing the emergence of a type-III Dirac cone combining flat and dispersive bands. **d** Measured dispersion along the $k_x$ direction for $k_y = -0.5 \times 2\pi/3d$ (line 2 in **b**), crossing the Dirac point in the dashed rectangle in **c**. **e** Dispersion obtained from the tight-binding model zoomed in the momentum space region marked with a dashed rectangle in **c** and **d**. In **a**, **c**, **d**, the white lines show the tight-binding model.

## Topological invariants of tilted and type-III Dirac cones

The emergence of unconventional Dirac cones when applying a uniaxial deformation to the orbital lattice can be understood from topological arguments. The analysis of the topological charge associated to each cone, that is, the winding of the Hamiltonian around each Dirac point, provides a clear picture of their birth and evolution. The vicinity of the band touching point around the Γ point for $\beta = 1$ can be described by an effective $2 \times 2$ Hamiltonian obtained from the projection of the full $4 \times 4$ Hamiltonian on the subspace of the two lowest energy bands[38], resulting in the following effective Hamiltonian:

$$H(\boldsymbol{q}) = -\frac{3}{2}t_L\sigma_0 + \frac{3}{8}t_L\begin{pmatrix} q_x^2 & q_xq_y \\ q_xq_y & q_y^2 \end{pmatrix} \equiv -\frac{3}{2}t_L\sigma_0 + \boldsymbol{h}(\boldsymbol{q}) \cdot \boldsymbol{\sigma}. \tag{2}$$

Matrix elements coupling the high and low energy manifolds are treated in a second order perturbation theory[38]. In this form, the winding of $\boldsymbol{h}(\boldsymbol{q})$ can be calculated as a winding vector[36]: $\boldsymbol{\mathcal{W}} = (2\pi)^{-1}\oint \boldsymbol{n}(\boldsymbol{q}) \times d\boldsymbol{n}(\boldsymbol{q})$, where $\boldsymbol{n}(\boldsymbol{q}) = \frac{\boldsymbol{h}(\boldsymbol{q})}{|\boldsymbol{h}(\boldsymbol{q})|}$, and the integral is performed along a closed line in momentum space that encircles the band touching point (see Supplementary Information). The modulus of $\boldsymbol{\mathcal{W}}$ is always an integer, thus providing a topological charge to the touching point, and it has a value $\mathcal{W} = 2$ in this case.

We can extend this analysis to the emerging Dirac cones when $\beta \neq 1$. Following the same procedure, the tight-binding Hamiltonian can be reduced to an effective $2 \times 2$ matrix close to the considered Dirac cone. For $\beta < 1$, the energy of the Dirac cones in the lower part of the spectrum is $E_D = E_0 - \frac{1}{2}\sqrt{3 + 6\beta^2} t_L$. Taking $E_D$ as the origin of energies, the effective Hamiltonian near a Dirac point reads:

$$H(\boldsymbol{q}) = v \cos\theta \, q_x \sigma_0 + v q_x \sigma_\theta + v \cos\theta \, q_y \sigma_x, \qquad (3)$$

where $v = \frac{\sqrt{3}}{4}\sqrt{|1-\beta^2|} dt_L$, the angle $\theta \in [0, \pi/2]$ is defined as $\tan\theta = \sqrt{1-\beta^2}/\sqrt{3}\beta$ (with $\theta \to \pi - \theta$ for the other Dirac point), and $\sigma_\theta = \boldsymbol{\sigma} \cdot \boldsymbol{u_\theta}$, with $\boldsymbol{u_\theta} = -\sin\theta \, \boldsymbol{u_y} + \cos\theta \, \boldsymbol{u_z}$, where $\boldsymbol{u_{x,y,z}}$ are Cartesian unit vectors. Analogously, for $\beta > 1$, around a type-III Dirac point ($E_D = E_0 - \frac{3}{2} t_L$), the reduced Hamiltonian reads:

$$H(\boldsymbol{q}) = v \cos\phi \, q_y \sigma_0 + v q_x \sigma_\phi - v \cos\phi \, q_y \sigma_z, \qquad (4)$$

where $\tan\phi = \sqrt{\beta^2 - 1}/\sqrt{4 - \beta^2}$ and $\sigma_\phi = \boldsymbol{\sigma} \cdot \boldsymbol{u_\phi}$, with $\boldsymbol{u_\phi} = \cos\phi \, \boldsymbol{u_x} - \sin\phi \, \boldsymbol{u_y}$ (for the other Dirac point: $\phi \to \pi - \phi$). By comparing $H(\boldsymbol{q})$ in Eqs. 3 and 4 with the generalised Dirac Hamiltonian, we can directly extract the Dirac effective velocities as a function of $\beta$ (shown as solid lines in Fig. 3f for $v_{0x}$ and $v_x$) as well as the tilt parameter $\tilde{v}_0$:

|  | $|v_{0x}|$ | $|v_{0y}|$ | $|v_x|$ | $|v_y|$ | $\tilde{v}_0$ |
|---|---|---|---|---|---|
| $\beta < 1$ | $v \cos\theta$ | 0 | $v$ | $v \cos\theta$ | $\cos\theta$ |
| $\beta > 1$ | 0 | $v \cos\phi$ | $v$ | $v \cos\phi$ | 1 |

To compute the winding of the Dirac cones, Hamiltonians 3 and 4 can be rearranged similarly to the rightmost hand side of Eq. 2, with a term in the form $\boldsymbol{h}(\boldsymbol{q}) \cdot \boldsymbol{\sigma}$. For $\beta < 1$ (Hamiltonian 3), the winding of $\boldsymbol{h}(\boldsymbol{q})$ around each of the Dirac cones is $\mathcal{W} = 1$, as indicated in the right panels of Fig. 3. This is also the case for the type-III Dirac cones for $\beta > 1$ (winding of Hamiltonian 4). A prominent feature of these Hamiltonians is that the vector $\boldsymbol{h}(\boldsymbol{q})$ winds on a plane that depends on the deformation $\beta$, namely on the plane $(\boldsymbol{u_x}, \boldsymbol{u_\theta})$ for $\beta < 1$, and $(\boldsymbol{u_\phi}, \boldsymbol{u_z})$ for $\beta > 1$ (see Supplementary Information).

The analysis of the windings sheds light on the mechanisms behind the creation of Dirac cones starting from a flat band touching a dispersive band. The single band touching at $\beta = 1$ is described by a winding $\mathcal{W} = 2$. When $\beta \lessgtr 1$ the band touching evolves into a pair of Dirac cones (tilted or type-III) with winding $\mathcal{W} = 1$. This is an illustration of one of the two possible scenarios for Dirac merging in two dimension, characteristic of, for example, bilayer graphene[37]. Remarkably, the orbital p-bands show the two possible universal scenarios for the merging of Dirac cones in two dimensions[39]: (i) two Dirac cones with $\mathcal{W} = 1$ emerge from a point with $\mathcal{W} = 2$, observed in the lower part of the spectra of Fig. 3; (ii) two Dirac cones with opposite winding $\mathcal{W} = \pm 1$ merge in a semi-Dirac cone with $\mathcal{W} = 0$, reported in the central part of the spectra in Fig. 3 (a $2 \times 2$ effective Hamiltonian analysis can also be done for the Dirac points at $E_0$).

The photonic realisation here reported demonstrates the flexibility of orbital bands to implement unconventional Dirac points. This is an asset for the engineering of photonic materials that combine different types of Dirac dispersions, an promising configuration for the study of analogue black holes in photonics[21,40]. Moreover, our experiments provide a recipe for the implementation of Dirac cones in solid state materials: the touching of a flat and a dispersive band with winding $\mathcal{W} = 2$ evolves into two Dirac cones in the presence of strain. This behaviour has been predicted for other lattice geometries[36], and it presents a natural playground to investigate the transition between different topological phases when particle interactions are present[41,42] or when time reversal symmetry is broken. Polaritons are particularly well suited to study these scenarios: thanks to their excitonic component they present significant repulsive interactions in the high density regime[43] and they are sensitive to external magnetic fields, allowing the implementation of quantum Hall phases[44,45]. Polariton orbital bands of the kind here reported open exciting perspectives for the study of topological lasers[46] and of Dirac superfluids[47].

**Acknowledgements**. This work was supported by the ERC grant Honeypol, the EU-FET Proactive grant AQuS, the French National Research Agency (ANR) project Quantum Fluids of Light (ANR-16-CE30-0021) and the Labex CEMPI (ANR-11-LABX-0007) and NanoSaclay (ICQOQS, Grant No. ANR-10-LABX-0035), the French RENATECH network, the CPER Photonics for Society P4S, and the Métropole Européenne de Lille via the project TFlight. T.O. acknowledges support from the Interdisciplinary Theoretical and Mathematical Sciences Program (iTHEMS) at RIKEN

# Supplementary Information

# Tilted and type-III Dirac cones emerging from flat bands in photonic orbital graphene

M. Milicevic[1], G. Montambaux[2], T. Ozawa[3], I. Sagnes[1], A. Lemaître[1], L. Le Gratiet[1], A. Harouri[1], J. Bloch[1], A. Amo[4]

[1]Centre de Nanosciences et de Nanotechnologies, CNRS, Univ. Paris-Sud, Université Paris-Saclay, C2N-Marcoussis, 91460 Marcoussis, France

[2]Laboratoire de Physique des Solides, CNRS, Univ. Paris-Sud, Université Paris-Saclay, 91405 Orsay Cedex, France

[3] Interdisciplinary Theoretical and Mathematical Sciences Program (iTHEMS), RIKEN, Wako, Saitama 351-0198, Japan

[4]Univ. Lille, CNRS, UMR 8523 – PhLAM – Physique des Lasers Atomes et Molécules, F-59000 Lille, France

Content :

**I.- Sample description and experimental set-up**

The sample used in the experiments is a semiconductor microcavity grown by molecular beam epitaxy. Two Bragg mirrors made of 28 (top) and 40 (bottom) pairs of $\lambda/4$ alternating layers of $Ga_{0.05}Al_{0.95}As/Ga_{0.80}Al_{0.20}As$ embed a $\lambda/2$ $Ga_{0.80}Al_{0.20}As$ cavity. $\lambda = 775$ nm is the emission/absorption wavelength of free excitons of 12 GaAs, 7 nm wide quantum wells grown in groups of four in the three central maxima of the electromagnetic field of the cavity. At 10 K, the temperature of the experiments, the excitonic and photonic resonances are in the strong coupling regime, with a Rabi splitting of 15 meV. The experiments are performed at a photon-exciton detuning of -10 meV (measured at the energy of the lowest energy flat band), except for the data shown in Fig. 3c, for which the detuning is -2 meV.

To fabricate the honeycomb lattices, the as-grown planar structure is subject to e-beam lithography and Inductively Coupled Plasma etching down to the GaAs substrate. Each micropillar in the lattice has a diameter of 2.75 μm, and the centre-to-centre distance varies between 2.4 and 2.72 μm, ensuring the overlap between adjacent micropillars. The polariton lifetime measured in a similar unetched microcavity is of the order of 30 ps. In the etched structures, the lateral defects induced during the microfabrication process reduce the lifetime to about 3-5 ps.

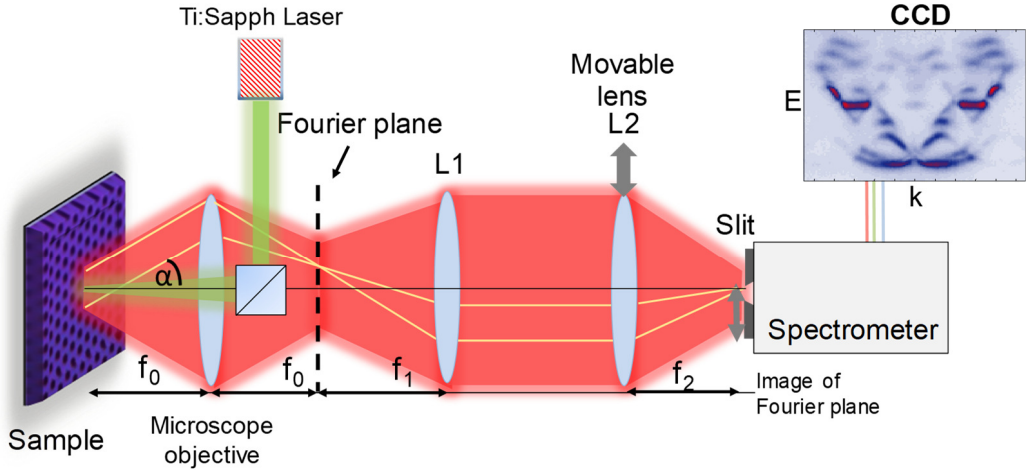

**Figure S1. Experimental setup.** Sketch of the experimental set-up employed to measure the momentum and energy resolved photoluminescence. Lenses L1 and L2 image the Fourier plane of the microscope objective on the entrance slit of the spectrometer, resulting in a $k_x$ vs energy image for a fixed value of $k_y$. By moving laterally lens L2 different values of $k_y$ can be accessed.

Experiments are performed by exciting the centre of each lattice with a continuous wave laser focalised in a spot of 3 μm in diameter (full width at half maximum). The wavelength of the laser is 745 nm (1.6642 eV), corresponding to an energy about 100 meV above the lower polariton s- and p-bands. The laser injects electron and holes that relax down incoherently to form polaritons that populate the lower bands of the lattice. The photons that escape out of the sample conserve the energy and in-plane momentum of polaritons in the lattice. The latter is related to the angle of emission following the expression: $k_\parallel = \frac{E}{\hbar c} sin(\alpha)$, where $\alpha$ is the emission angle measured from the normal to the lattice plane, $E$ is the energy of the emitted photon and $c$ is the speed of light in vacuum. Therefore, by resolving the emission in energy and in angle we can measure the polariton dispersion.

Figure S1 shows the experimental set-up, in which the Fourier plane of the excitation/collection lens (microscope objective) is imaged on the entrance slit of a spectrometer attached to a CCD. We select the emission linearly polarized along the $k_y$ direction, parallel to the grooves of the grating in the spectrometer. Similar results were obtained for the perpendicular polarisation.

A single image of the CCD registers the emission energy as a function of $k$ parallel to the entrance slit (as shown in the figure), for a fixed value of $k$ perpendicular to the entrance slit of the spectrometer. By taking images for different lateral positions of lens L2, we can perform a full tomography of the three-dimensional space "energy, $k_x, k_y$". Post-processing of the data allows displaying energy-momentum cuts along any momentum direction or momentum-momentum cuts for a given emission energy.

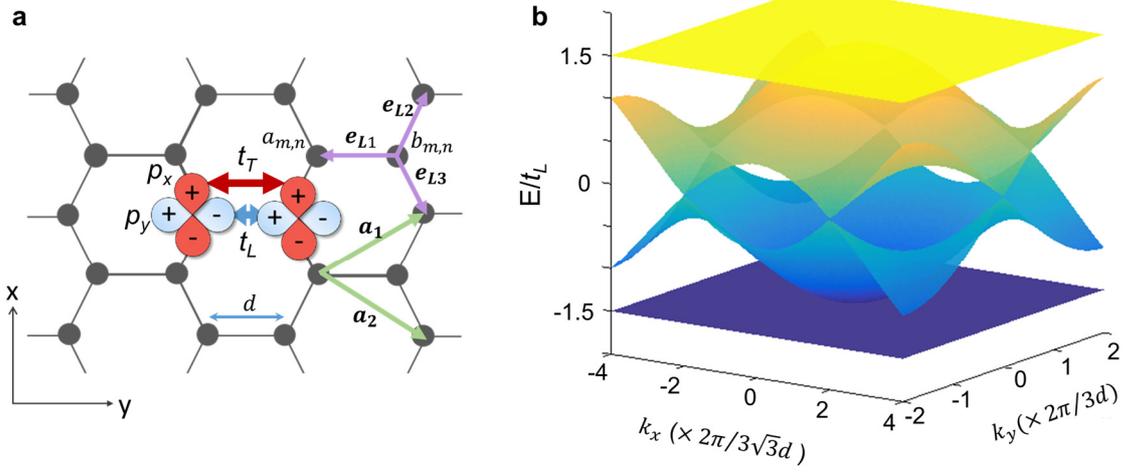

**Figure S2. Tight-binding dispersions. a** Scheme of the honeycomb lattice and the unit vectors used in the tight binding model. **b** Spectrum of the p-bands obtained from the diagonalization of Hamiltonian E2 with $t_T = 0$ and $\beta = 1$.

II.- Tight-binding Hamiltonian and extraction of $\beta$

The coupling of the $p_x$, $p_y$ orbitals in the isotropic honeycomb lattice can be described by a tight-binding Hamiltonian of the form[1,2]:

$$H = -\sum_{m,n}\{t_L(\hat{a}^\dagger_{m,n} \cdot \boldsymbol{e}_{L1})(\boldsymbol{e}_{L1} \cdot \hat{b}_{m,n}) + t_L(\hat{a}^\dagger_{m+1,n} \cdot \boldsymbol{e}_{L2})(\boldsymbol{e}_{L2} \cdot \hat{b}_{m,n}) + t_L(\hat{a}^\dagger_{m,n+1} \cdot \boldsymbol{e}_{L3})(\boldsymbol{e}_{L3} \cdot \hat{b}_{m,n+1}) +$$

$$t_T(\hat{a}^\dagger_{m,n} \cdot \boldsymbol{e}_{T1})(\boldsymbol{e}_{T1} \cdot \hat{b}_{m,n}) + t_T(\hat{a}^\dagger_{m+1,n} \cdot \boldsymbol{e}_{T2})(\boldsymbol{e}_{T2} \cdot \hat{b}_{m,n}) + t_T(\hat{a}^\dagger_{m,n+1} \cdot \boldsymbol{e}_{T3})(\boldsymbol{e}_{T3} \cdot \hat{b}_{m,n+1}) + h.c.\},$$

(E1)

where, $\boldsymbol{e}_{L1,L2,L3}$ are unit vectors oriented parallel to the nearest-neighbour links of each site (see Fig. S2a), $\boldsymbol{e}_{T1,T2,T3}$ are unit vectors oriented perpendicular to $\boldsymbol{e}_{L1,L2,L3}$, respectively, $\hat{a}^\dagger_{m,n} = \left(\hat{a}^\dagger_{m,n\,(x)}, \hat{a}^\dagger_{m,n\,(y)}\right)$ and $\hat{b}^\dagger_{m,n} = \left(\hat{b}^\dagger_{m,n\,(x)}, \hat{b}^\dagger_{m,n\,(y)}\right)$, with $\hat{a}^\dagger_{m,n\,(x,y)}$, $\hat{b}^\dagger_{m,n\,(x,y)}$ the creation operators of a photon in the $p_x$ or $p_y$ orbital of the $a$, $b$ site of the $m, n$ unit cell in the lattice. If the hopping of orbitals oriented perpendicular to the links is negligible ($t_T = 0$), Hamiltonian E1 can be written in momentum space in the form[2]:

$$H(\boldsymbol{k}) = -t_L \begin{pmatrix} 0_{2\times 2} & Q(\boldsymbol{k}) \\ Q(\boldsymbol{k})^\dagger & 0_{2\times 2} \end{pmatrix}, \text{ with } Q(\boldsymbol{k}) = \begin{pmatrix} f_1 & g \\ g & f_2 \end{pmatrix},$$

(E2)

where $f_1 = \frac{3}{4}(e^{i\boldsymbol{k}\cdot\boldsymbol{a}_1} + e^{i\boldsymbol{k}\cdot\boldsymbol{a}_2})$, $f_2 = 1 + \frac{1}{4}(e^{i\boldsymbol{k}\cdot\boldsymbol{a}_1} + e^{i\boldsymbol{k}\cdot\boldsymbol{a}_2})$, and $g = \frac{\sqrt{3}}{4}(e^{i\boldsymbol{k}\cdot\boldsymbol{a}_1} - e^{i\boldsymbol{k}\cdot\boldsymbol{a}_2})$, with $\boldsymbol{a}_1 = (\frac{\sqrt{3}}{2}d, \frac{3}{2}d)$, $\boldsymbol{a}_2 = (-\frac{\sqrt{3}}{2}d, \frac{3}{2}d)$ the lattice vectors depicted in Fig. S2a and $\boldsymbol{k} = (k_x, k_y)$ the quasi-momentum. The diagonalization of Hamiltonian E2 results in four bands depicted in Fig. S2b. The spectrum shows two flat bands at high and low energy that touch two dispersive bands with six Dirac points (only two are non-equivalent).

If we account for different hoppings $t_L$ and $t'_L$ for different links, Hamiltonian E2 is still valid with the same expressions for $f_1$ and $g$, and $f_2$ replaced by $f_2 = \beta + \frac{1}{4}(e^{i\mathbf{k}\cdot\mathbf{a}_1} + e^{i\mathbf{k}\cdot\mathbf{a}_2})$. As a function of $\beta = t'_L/t_L$, its spectrum $\varepsilon(\mathbf{k})$ takes the form:

$$\varepsilon(\mathbf{k}) = \pm\sqrt{E_0(\mathbf{k}) \pm E(\mathbf{k})} \qquad (E3)$$

where the four sign combinations make the four bands and:

$$E(\mathbf{k}) = t_L \left[\frac{3}{4}\sin^2(K_x) \cdot \left(\beta^2 + \cos^2(K_x) - 2\beta\cos(K_x)\cos(K_y)\right) + \right.$$
$$\left. \frac{1}{4}\left(2\cos^2(K_x) - \beta^2 - \beta\cos(K_x)\cos(K_y)\right)^2\right]^{1/2}, \qquad (E4)$$

$$E_0(\mathbf{k}) = \frac{t_L}{4}\left(3 + 2\beta^2 + 2\cos^2(K_x) + 2\beta\cos(K_x)\cos(K_y)\right), \qquad (E5)$$

with $K_x \equiv \frac{\sqrt{3}}{2}k_x d$ and $K_y \equiv \frac{3}{2}k_y d$.

For $\beta < 1$ the energy $E_D$ and momentum space coordinates of the low energy emerging Dirac cones are:

$$E_D = E_0 - \frac{1}{2}\sqrt{3 + 6\beta^2}\, t_L, \; K_y = 0, \; \cos K_x = \beta, \qquad (E6a)$$

while for $\beta > 1$ they are:

$$E_D = E_0 - \frac{3}{2}t_L, \; \cos K_y = \frac{2-\beta^2}{\beta} 0, \; K_x = 0, \qquad (E6b)$$

In Figs. 3 and 4 of the main text, the experimental value of $\beta$ is obtained from fits of the spectrum of the tight-binding model to the measured dispersions, in which $t_T = 0$ and $t_L$, $\beta$ and the onsite energy $E_0$ are the fitting parameters.

## III.- Gap opening at $\beta = 0.45$

Figure 3c of the main text shows the formation of a semi-Dirac cone at $E = E_0$ for $\beta = 0.5$. If $\beta$ is further reduced to 0.45, a gap opens. To further evidence the gap opening, Fig. S3 shows the spectrum obtained from Figs. 3c and d at the momentum of the Dirac point ($k_x = -3 \times 2\pi/3\sqrt{3}d$) for $\beta = 0.5$ and $\beta = 0.45$. In the latter case a splitting is apparent at $E_0$, corresponding to the gap opening.

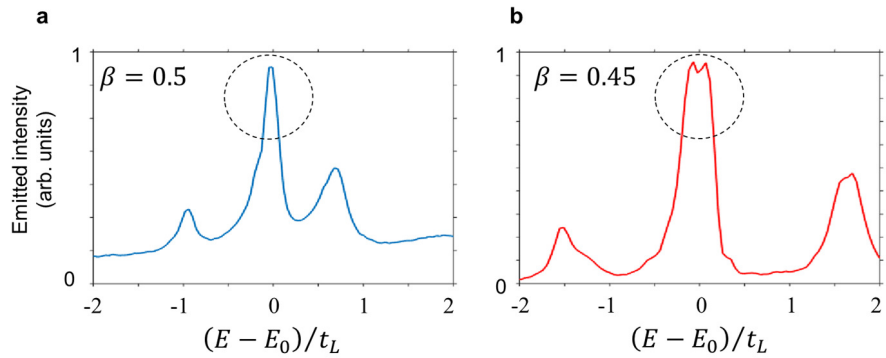

**Figure S3. Spectrum at gap opening. a**, **b** Photoluminescence spectra taken at the momentum position of the semi-Dirac cone $k_x = -3 \times 2\pi/3\sqrt{3}d$, $k_y = 2\pi/(d' + d/2)$, for $\beta = 0.5$ (**a**) and $\beta = 0.45$ (**b**), corresponding to Fig. 3c and d, respectively. In **b**, a splitting is apparent in the circled area.

## IV.- Tomography of tilted Dirac cones

Figure 3 of the main text shows the emergence of tilted Dirac cones from the flat band – dispersive band touching when $\beta < 1$. The tilt is clearly visible in the $k_x$ direction, resulting in a non-zero value of $v_{0x}$ (as defined in Eq. 1). Figure S4 shows the dispersion across one of the newly created Dirac cones in the $k_y$ direction for $\beta = 0.5$. The white solid lines depict the eigenvalues calculated by diagonalising Hamiltonian E2. As it is apparent from the measured spectra and the tight-binding bands, the newly created cones are not tilted along the $k_y$ direction, resulting in $v_{0y} \approx 0$.

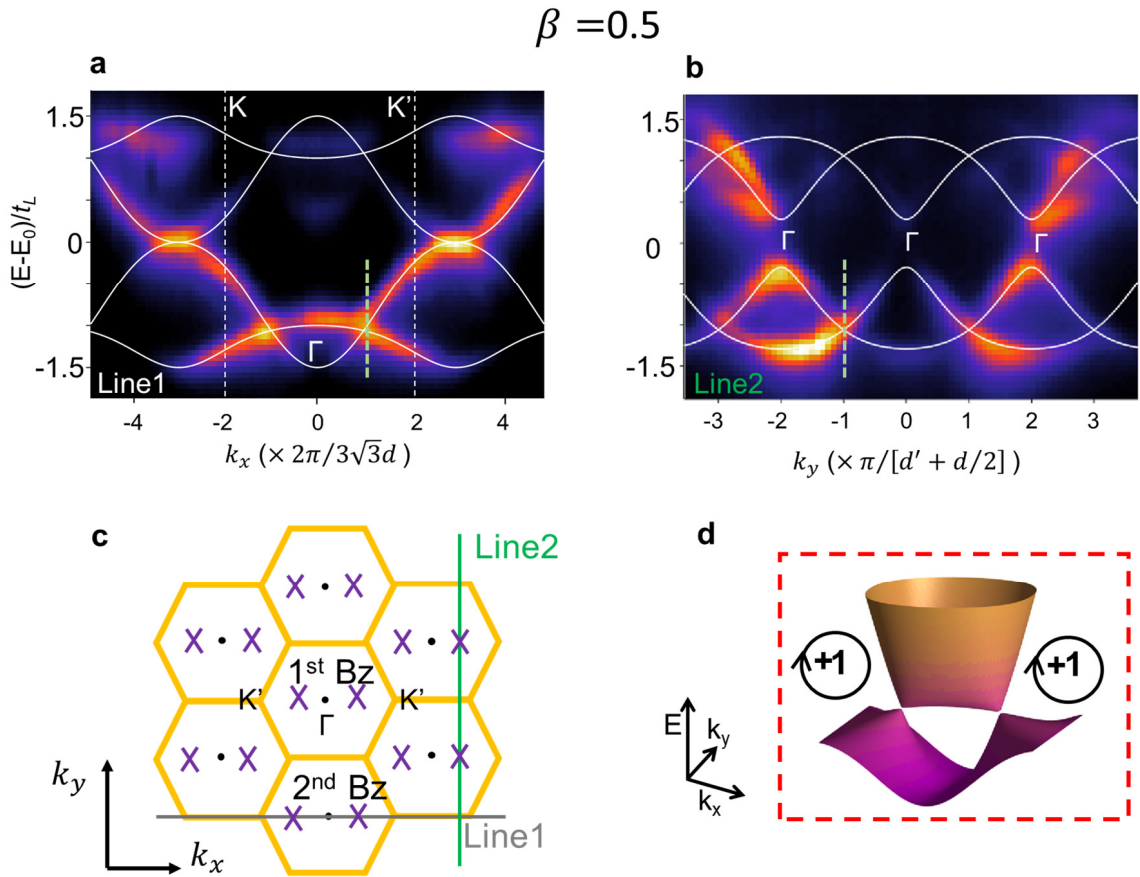

**Figure S4. Tomography of tilted Dirac cones. a** Measured dispersion along the $k_x$ direction for $k_y = 4\pi/(d' + d/2)$ (line 1 in panel **c**) for $\beta = 0.5$ (same as Fig. 3c). **b** Measured dispersion along the $k_y$ direction for $k_x = 2\pi/3\sqrt{3}d$, traversing the tilted Dirac cone (green dashed line in **a**, line 2 in **c**) . **c** Sketch of the different Brillouin zones in momentum space. The orange hexagons show the Brillouin zones as defined for $\beta = 1$. The crosses indicate the position of the emergent tilted Dirac cones. **d** Zoom of the tight-binding eigenenergies in the region close to the lowest energy Dirac points. Parameters are the same as for Fig. 3c.

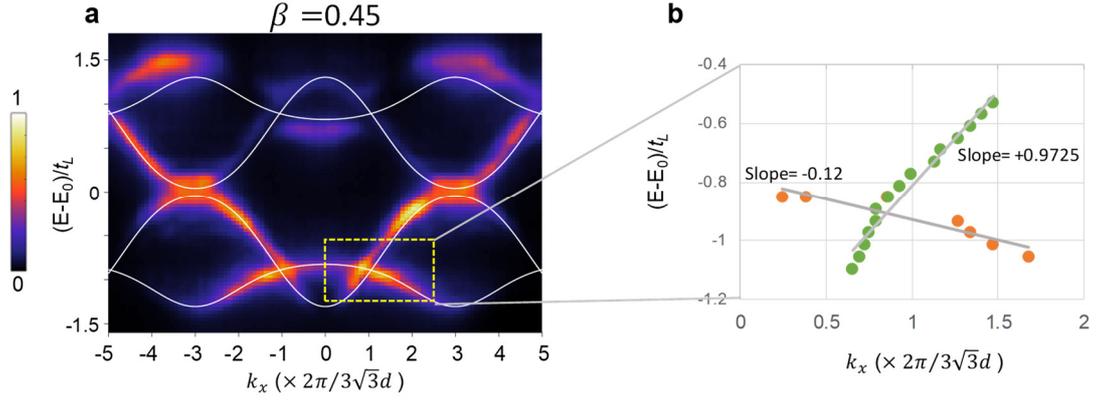

**Figure S5. a** Measured dispersion along the $k_x$ direction for $k_y = 2\pi/(d' + d/2)$ for $\beta = 0.45$ (same as Fig. 3d). **b** Extracted data points from the Dirac point marked in a dashed rectangle in **a**. From linear fits to the points we obtain $|v_{0x}|$ and $|v_x|$, plotted in Fig. 3f of the main text.

## V.- Measurement and calculation of the effective Dirac velocities

Figure 3f of the main text shows the effective Dirac velocities $v_{0x}$ and $v_x$ measured from the measured dispersions along the $k_x$ direction plotted in Fig. 3. To measure these velocities, we first extract the dispersion around the Dirac cones from the maximum photoluminescence intensity as a function of energy and momentum. For the case of $\beta = 0.45$, the data points are shown in Fig. S5b. A linear fit allows measuring the slope of the dispersion. By comparison with Eq. 1 of the main text, the two measured slopes correspond to $|v_{0x}| - |v_x|$ and $|v_{0x}| + |v_x|$. From that, we extract $|v_{0x}|$ and $|v_x|$. We repeat the procedure for the symmetric Dirac point and we take the average of their absolute value.

Note that for type-I Dirac cones, $v_{0x} = v_{0y} = 0$, and the absolute value of the two slopes should be identical and equal to $v_x$. To test this hypothesis, we have measured the slope around the K' type-I Dirac cone at $E_0$ in Fig. 3a at positive momenta. After converting the slopes to effective Weyl velocities (Eq. 1 of the main text) we obtain $|v_{0x}| = 0.07 \pm 0.03$ and $|v_x| = 0.72 \pm 0.03$ in units of $dt_L$.

From the tight-binding Hamiltonian, the effective Dirac velocities can be obtained from the derivative along $k_x$ and $k_y$ of the eigenvalues of E3-E5 at the position of the emergent Dirac cones (E6), resulting in the analytical expressions for $\beta < 1$ (normalised to $dt_L$, where we have assumed that in the experiment $d \approx d'$):

$$|v_{0x}| = \frac{3\beta}{4}\sqrt{\frac{1-\beta^2}{1+2\beta^2}}, \qquad |v_{0y}| = 0, \qquad |v_x| = \frac{\sqrt{3}}{4}\sqrt{1-\beta^2}, \qquad |v_y| = \frac{\sqrt{3-3\beta^2}}{4}. \tag{E7}$$

The analytical expressions for $|v_{0x}|$ and $|v_x|$ are shown in solid lines in Figs. S5 and 3f.

For $1 < \beta < 2$ we have (same units as above):

$$|v_{0x}| = 0, \qquad |v_{0y}| = |v_y| = \frac{1}{4}\sqrt{(\beta^2-1)(4-\beta^2)}, \qquad |v_x| = \frac{\sqrt{3}}{4}\sqrt{\beta^2-1}. \tag{E8}$$

Note that the fact that $|v_{0y}| = |v_y|$ (with opposite sign: $v_{0y} = -v_y$) for $\beta > 1$ is the signature of the type-III Dirac cone, and it results in a flat band along the $k_y$ direction.

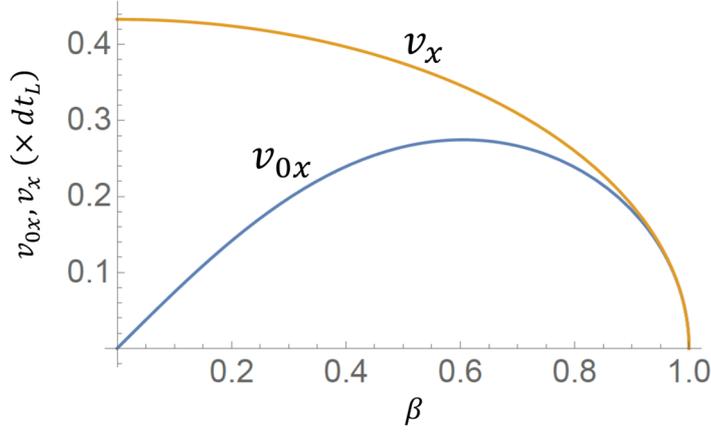

**Figure S5.** $v_{0x}$ and $v_x$ as a function of $\beta$ calculated from the derivative of the tight-binding spectrum at the position of the emergent Dirac cones.

## VI.- Hamiltonian reduction and winding around the band touching points

The vicinity of touching point between two bands can be described by an effective $2 \times 2$ Hamiltonian obtained from the projection of the full $4 \times 4$ Hamiltonian on the subspace of the two lowest energy bands[3]. The reduced Hamiltonian can be expressed in the generic form:

$$H(\boldsymbol{q}) = \alpha(\boldsymbol{q})\sigma_0 + \boldsymbol{h}(\boldsymbol{q}) \cdot \boldsymbol{\sigma} = \alpha(\boldsymbol{q})\sigma_0 + h_a(\boldsymbol{q}) \cdot \sigma_a + h_b(\boldsymbol{q}) \cdot \sigma_b, \tag{E9}$$

where $\sigma_0$ is the $2 \times 2$ identity matrix and $\sigma_{a,b} = \boldsymbol{u}_{a,b} \cdot \boldsymbol{\sigma}$, with $\boldsymbol{u}_{a,b}$ unit vectors and $\boldsymbol{\sigma} = (\sigma_x, \sigma_y, \sigma_z)$ is the vector of Pauli matrices. Without loss of generality and for simplicity, we consider the case of $\boldsymbol{u}_a \cdot \boldsymbol{u}_b = 0$, which implies $\{\sigma_a, \sigma_b\} = 0$.

The winding vector of this Hamiltonian around the band touching point is[4]:

$$\boldsymbol{W} = \frac{1}{2\pi}\oint \boldsymbol{n}(\boldsymbol{q}) \times d\boldsymbol{n}(\boldsymbol{q}) = \frac{\boldsymbol{u}_c}{2\pi}\oint \frac{h_a(\boldsymbol{q})\nabla h_b(\boldsymbol{q}) - h_b(\boldsymbol{q})\nabla h_a(\boldsymbol{q})}{h_a^2(\boldsymbol{q}) + h_b^2(\boldsymbol{q})} d\boldsymbol{q}, \tag{E10}$$

where $\boldsymbol{u}_c = \boldsymbol{u}_a \times \boldsymbol{u}_b$ and $\boldsymbol{n}(\boldsymbol{q}) = \frac{\boldsymbol{h}(\boldsymbol{q})}{|\boldsymbol{h}(\boldsymbol{q})|}$. The integral is performed along a closed line in momentum space that encircles the band touching point. The modulus of $\boldsymbol{W}$ can only take integer values.

The effective $2 \times 2$ Hamiltonian described in the main text close to the flat-parabolic touching at the $\Gamma$ point for $\beta = 1$ takes the form:

$$H(\boldsymbol{q}) = -\frac{3}{2}t_L\sigma_0 + \frac{3}{8}t_L\begin{pmatrix} q_x^2 & q_xq_y \\ q_xq_y & q_y^2 \end{pmatrix} \equiv -\frac{3}{2}t_L\sigma_0 + \boldsymbol{h}(\boldsymbol{q}) \cdot \boldsymbol{\sigma}, \tag{E11}$$

and it is characterized by the winding vector $\boldsymbol{W} = +2\,\boldsymbol{u}_z$.

For $\beta < 1$, close to the emerging Dirac cone, it reads:

$$H(\boldsymbol{q}) = v\cos\theta\, q_x\sigma_0 + vq_x\sigma_\theta + v\cos\theta\, q_y\sigma_x, \tag{E12}$$

where $v = \frac{\sqrt{3}}{4}\sqrt{|1-\beta^2|}dt_L$, the angle $\theta \in [0,\,\pi/2]$ is defined as $\tan\theta = \sqrt{1-\beta^2}/\sqrt{3}\beta$ (with $\theta \to \pi - \theta$ for the other Dirac cone), and $\sigma_\theta = \boldsymbol{\sigma}\cdot\boldsymbol{u_\theta}$, with $\boldsymbol{u_\theta} = -\sin\theta\,\boldsymbol{u_y} + \cos\theta\,\boldsymbol{u_z}$, where $\boldsymbol{u_{x,y,z}}$ are Cartesian unit vectors.

Analogously, for $\beta > 1$, around one of the type-III Dirac points, the reduced Hamiltonian reads:

$$H(\boldsymbol{q}) = v\cos\phi\, q_y\sigma_0 + vq_x\sigma_\phi - v\cos\phi\, q_y\sigma_z, \tag{E13}$$

where $\tan\phi = \sqrt{\beta^2 - 1}/\sqrt{4-\beta^2}$ and $\sigma_\phi = \boldsymbol{\sigma}\cdot\boldsymbol{u_\phi}$, with $\boldsymbol{u_\phi} = \cos\phi\,\boldsymbol{u_x} - \sin\phi\,\boldsymbol{u_y}$ (for the other Dirac point: $\phi \to \pi - \phi$).

By comparing the form of Eqs. E12 and E13 with the generalised Dirac Hamiltonian presented in the main text and its eigenvalues in Eq. 1, we can write the effective velocities and tilt parameter $\tilde{v}_0$ as a function of $v$, $\theta$ and $\phi$:

|  | $|v_{0x}|$ | $|v_{0y}|$ | $|v_x|$ | $|v_y|$ | $\tilde{v}_0 = \sqrt{\left(\frac{v_{0x}}{v_x}\right)^2 + \left(\frac{v_{0y}}{v_y}\right)^2}$ |
|---|---|---|---|---|---|
| $\beta < 1$ | $v\cos\theta$ | $0$ | $v$ | $v\cos\theta$ | $\cos\theta$ |
| $\beta > 1$ | $0$ | $v\cos\phi$ | $v$ | $v\cos\phi$ | $1$ |

If $v$, $\theta$ and $\phi$ are explicitly expressed in terms of $\beta$ we obtain Eqs. E7 and E9, in agreement with the full tight-binding calculation.

The modulus of the winding vector for Hamiltonian E11 is $\mathcal{W} = 2$, while for each of the tilted ($\beta < 1$) and type-III Dirac cones ($\beta > 1$), the winding is $\mathcal{W} = 1$. As stated in the main text, a prominent feature of these Hamiltonians is that vector $\boldsymbol{h}(\boldsymbol{q})$ winds on a plane whose orientation depends on the deformation $\beta$. Namely, for $\beta < 1$ the pseudo-field $\boldsymbol{h}(\boldsymbol{q})$ resides on the plane $(\boldsymbol{u_x}, \boldsymbol{u_\theta})$ and the winding vector points in the direction $\frac{\boldsymbol{w}}{|\boldsymbol{w}|} = (0, \cos\theta, \pm\sin\theta)$. The winding vector plane thus rotates as a function of $\beta$. As each of the two emergent Diract cones is described by an angle $\theta$ of opposite sign, the winding plane turns in opposite directions for each Dirac cone when $\beta$ goes to zero. This is illustrated by the red and blue circles in Fig. S6, which show, respectively, the winding plane of each of the two Dirac cones.

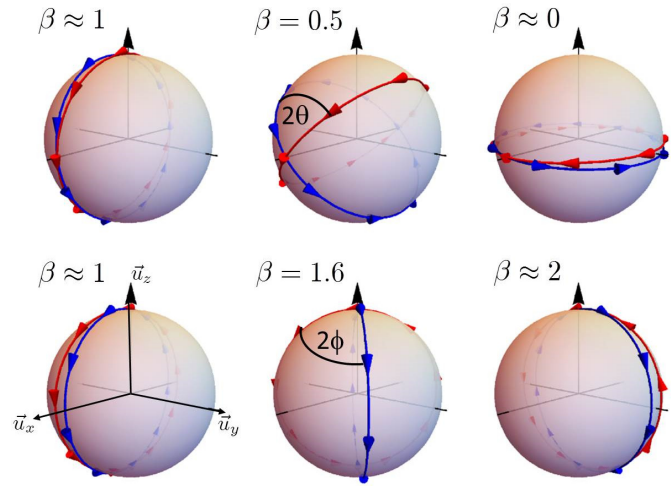

**Figure S6.** Scheme of the rotation of the winding plane as a function of $\beta$ for the two Dirac cones (red and blue lines) emerging from the parabolic-flat band touching.

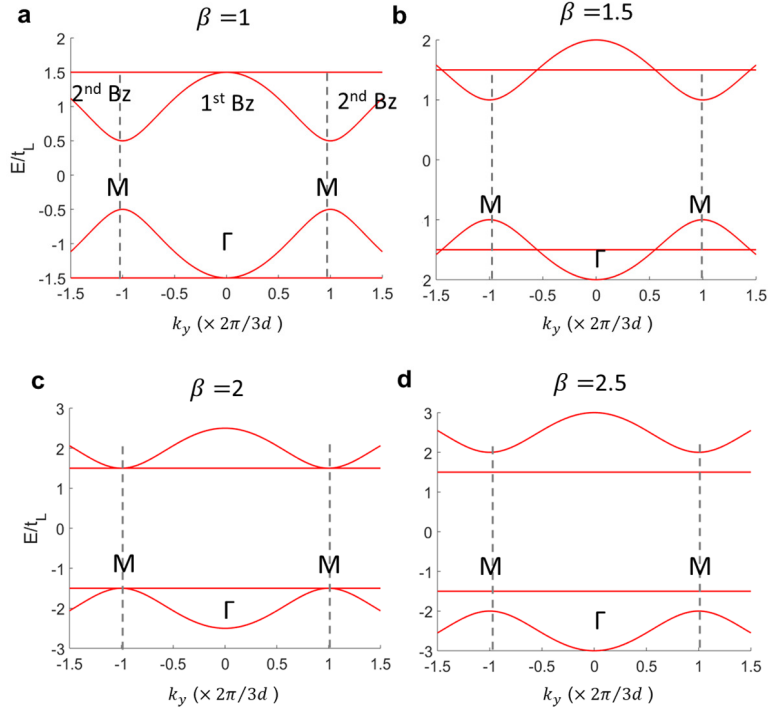

**Figure S7.** Tight binding spectra for different values of $\beta > 1$ at $k_x = 0$. At $\beta = 2$ the type-III Dirac cones merge and annihilate resulting in a gap opening for $\beta > 2$.

For $\beta > 1$, the vector $\boldsymbol{h}(\boldsymbol{q})$ resides on the plane $(\boldsymbol{u}_\phi, \boldsymbol{u}_z)$ and the winding vector points in the direction $\frac{\boldsymbol{w}}{|\boldsymbol{w}|} = (\pm \sin\phi, \cos\phi, 0)$. Note that at $\beta = 2$ (bottom-right panel in Fig. S6), the two emergent Dirac cones are described by a winding vector residing in the same plane but pointing in opposite directions. Therefore, the winding around each Dirac cone has opposite sign: $\mathcal{W} = +1$ and $\mathcal{W} = -1$. At $\beta = 2$, the Dirac cones merge at the M-point and annihilate, and a gap opens for $\beta > 2$. This situation is shown in Fig. S7, and it could not be observed experimentally because the engineered hoppings in the designed structure were limited to $\beta = 1.5$. This kind of merging preceded by a rotation of the winding plane has been recently discussed in detail in the context of the Mielke lattice under strain[4].